\documentstyle[12pt]{article}

\begin{document}

\sloppy

\begin{flushright}{UT-746\\ March '96}\end{flushright}

\vskip 1.5 truecm

\centerline{\large{\bf Alternative Approach to}}
\centerline{\large{\bf  Gaugino Condensation}}
\vskip .75 truecm
\centerline{\bf Tomohiro Matsuda
\footnote{matsuda@danjuro.phys.s.u-tokyo.ac.jp}}
\vskip .4 truecm
\centerline {\it Department of Physics, University of Tokyo}
\centerline {\it Bunkyo-ku, Tokyo 113,Japan}
\vskip 1. truecm

\makeatletter
\@addtoreset{equation}{section}
\def\theequation{\thesection.\arabic{equation}}
\makeatother

\vskip 1. truecm

\begin{abstract}
\hspace*{\parindent}
We examine the mechanism of  gaugino condensation in supersymmetric
theories  
within a Nambu-Jona-Lasinio type approach.
We investigate the effective Lagrangian description of higher
energy theories that include some moduli fields in the
gauge coupling constant.
First we consider supersymmetric QCD with and without a mass term.
We can find a phase transition in massless theory, but
when we add a mass term, such a phase transition disappears.
We also examine a model with a dilaton dependent coupling
and find that it is very similar to supersymmetric QCD.
Application of our method to supergravity is also examined.

\end{abstract}
\newpage

\section{Introduction}
\hspace*{\parindent}
There has recently been considerable attention focused on the study of 
supersymmetric models of elementary particle interactions.
This is especially true in the context of grand unification theories,
where remarkable studies have been done in the hope of solving the
gauge hierarchy problem or unifying the gravitational interaction
within the superstring formalism.
Supersymmetric extension of the gravity(supergravity) seems necessary 
in introducing soft breaking terms and making the cosmological
constant vanish simultaneously.
In supergravity models,  spontaneous breaking of local supersymmetry
or super-Higgs mechanism may generate soft supersymmetry breaking
terms that allow to fulfill such phenomenological requirements.
However, the super-Higgs mechanism implies the existence of a 
supergravity breaking scale, intermediate between the Planck scale($M_{p}$)
and the weak scale($M_{W}$).
The intermediate scale is expected to be of O($10^{13}$Gev).
Here we expect that this intermediate scale is implemented by the 
mechanism of gaugino condensation in the hidden sector which couples
to the visible sector by gravitational interactions.
The effective action for 
gaugino condensation is well studied by many authors\cite{rep,leff}.
The main purpose of this paper is to reproduce these results
 and seek for  new features in these theories 
by means of the Nambu-Jona-Lasinio method\cite{njlmat,ross}. 

In section 2 and 3 we consider an intermediate scale effective Lagrangian
for supersymmetric QCD with  $N_{c}$ colors and $N_{f}(<N_{c}-1)$ 
flavors and also consider the effect of the dilaton dependent
coupling constant. 
These theories have classical flat directions.
If matter fields develop expectation value $v$ on these directions,
the original $SU(N_{c})$ gauge symmetry is broken and 
the effective low energy
Lagrangian has two parts.
These are the kinetic terms for low energy pure 
supersymmetric Yang-Mills theory and
the Nambu-Goldstone field, and their interaction term of order O(1/$v$).
Taking $v$ to infinity, the resultant theory is a pure
Yang-Mills theory without any interaction, so we tend to think 
that even in the existence of higher interaction terms 
we can extend the analysis of pure Yang-Mills theory.
But this is merely a naive expectation so it seems important 
to analyze the theory from another point of view.
From this standpoint, using the Nambu-Jona-Lasinio method
we examine the effect of 
the higher terms that affect on gaugino condensation. 

Section 4 includes the extension to  supergravity.

\section{Gaugino condensation in supersymmetric QCD}
\hspace*{\parindent}
Gaugino condensation in supersymmetric gauge theories has been extensively
studied by many authors both in global\cite{rep} and 
local\cite{leff} theories.
In this section we examine the vacuum structures of 
Supersymmetric QCD(SQCD) theories with $N_{f}<N_{c}-1$
by using the Nambu-Jona-Lasinio method.
We follow ref.\cite{rep} in deriving the effective Lagrangian.

The results presented below nicely agree with the previous studies which are 
given by instanton or effective Lagrangian analysis.

Our starting point is a Lagrangian with a gauge group $SU(N_{c})$ with 
$N_{f}$ flavors of quarks.
These superfields can be written with component fields as:
\begin{equation}
  \label{component}
  \left\{
    \begin{array}{l}
      Q^{ir}=\phi^{ir}+\theta^{\alpha}\psi_{\alpha}^{ir}+\theta^{2}F^{ir}\\
      \overline{Q}_{ir}=\overline{\phi}_{ir}+\overline{\theta}_{\dot{\alpha}}
      \psi^{\dot{\alpha}}_{ir}+\overline{\theta}^{2}F_{ir}
    \end{array}
  \right.
\end{equation}
The gauge fields $A^{a}_{\mu}(a=1,...,N_{c}^{2}-1)$ are included in vector 
multiplets $V^{a}$ accompanied by their super-partners, gauginos $\lambda^{a}$
and auxiliary fields $D^{a}$.
The total theory is given by
\begin{eqnarray}
  \label{lag0}
  L&=&\frac{1}{4g^{2}}\int d^{2}\theta W^{\alpha a}W_{\alpha}^{a}+h.c.
  +\int d^{4}\theta \left[Q^{+}e^{V}Q+\overline{Q}e^{V}
  \overline{Q}^{+}\right]
\end{eqnarray}
Classically, this theory has a global $U(N_{f})_{Left}\times 
U(N_{f})_{Right}\times U(1)_{R}$ symmetry.
The $U(N_{f})_{Left}\times U(N_{f})_{Right}$ symmetry is just
like that of ordinary QCD, corresponding to separate rotation of the 
$Q$ and $\overline{Q}$ fields.
The symmetry $U(1)_{R}$ is a R-invariance, a symmetry under which 
the components of a given superfield transform differently.
This corresponds to a rotation of the phases of the grassmannian 
variables $\theta^{\alpha}$,
\begin{equation}
  \label{r}
  \left\{
  \begin{array}{lll}
    \lambda&\rightarrow&e^{i\alpha}\lambda\\
    \psi&\rightarrow&e^{i\alpha}\psi\\
    \overline{\psi}&\rightarrow&e^{i\alpha}\overline{\psi}
  \end{array}
  \right.
\end{equation}
or
\begin{equation}
  \left\{
    \begin{array}{lll}
      W_{\alpha}(\theta)&\rightarrow&e^{-i\alpha}W_{\alpha}(\theta 
               e^{i\alpha})\\
      Q(\theta)&\rightarrow&Q(\theta  e^{i\alpha})\\
      \overline{Q}(\theta)&\rightarrow&\overline{Q}(\theta  e^{i\alpha})
    \end{array}
  \right.
\end{equation}
Just as in ordinary QCD, some of these symmetries are explicitly 
broken by anomalies.
A simple computation shows that the following symmetry,
which is a combination of the ordinary chiral $U(1)$ and the $U(1)_{R}$
symmetry, is anomaly-free.
\begin{equation}
  \left\{
    \begin{array}{lll}
      W_{\alpha}(\theta)&\rightarrow&e^{-i\alpha}W_{\alpha}(\theta 
               e^{i\alpha})\\
      Q(\theta)&\rightarrow&e^{i\alpha(N_{c}-N_{f})/N_{f}}
               Q(\theta  e^{i\alpha})\\
      \overline{Q}(\theta)&\rightarrow&e^{i\alpha(N_{c}-N_{f})/N_{f}}
               \overline{Q}(\theta  e^{i\alpha})
    \end{array}
  \right.
\end{equation}
From now on, we call this non-anomalous global symmetry  $U(1)_{R'}$.

Since this model has flat directions, it is reasonable to
expect that $Q$ and $\overline{Q}$ may develop their vacuum
expectation values along these directions.
If $N_{f}<N-1$, the gauge group is not completely broken.
Moreover, we can see that instantons cannot generate a superpotential
in this case, so considering another type of non-perturbative effects 
in this model seems important.

For simplicity, here we consider the case: $SU(N_{c})$ gauge group is
broken to $SU(N_{c}-N_{f})$.
The low-energy theory consists of two parts: Kinetic terms
for the unbroken pure $SU(N_{c}-N_{f})$ 
gauge interaction and one for the 
massless chiral field.
In addition to these terms, we should include higher dimensional
operators.
A dimension-five operator, in general, is generated at 
one-loop level\cite{pol}.
This can be obtained also from the renormalization of the effective
coupling\cite{rep}:
\begin{equation}
  \label{geff}
  L=\frac{1}{4g^{2}}\left[1+\frac{g^{2}}{32\pi^{2}}N_{f}ln
  \left(\frac{\phi}{\Lambda}\right)\right]W^{\alpha}W_{\alpha}
\end{equation}
Of course, this term itself is not dimension five.
Redefining the field as $\phi=<\phi>+\phi'$, this term
produces a dimension five operator, namely $\sim\frac{\phi'}{<\phi>}W^{2}$.
$\phi$ must be chosen to be invariant under all non-R
symmetries.
Detailed arguments on such an field dependence of
coupling constant are given in ref.\cite{rep} and references therein.
The non-anomalous R'-symmetry of the original theory must be realized
in the effective low-energy Lagrangian by the shift
induced by $\phi$.
That determines the R'-charge of $\phi$ to be $(N_{c}-N_{f})/N_{f}$.

For simplicity, we consider a generalized form
\begin{equation}
  \label{1}
  L=\frac{1}{4}f(\phi)W^{\alpha}W_{\alpha}+h.c.
  +\phi^{*}\phi
\end{equation}
where $f(\phi)$ is the field dependent coupling constant.
\begin{eqnarray}
  \label{1a}
  f(\phi)&=&\frac{1}{g_{0}^{2}}+\beta log\left(\frac{\phi}{\Lambda}
  \right)\nonumber\\
\end{eqnarray}
Here $\beta$ is a constant chosen to realize the anomaly free
(mixed) R'-symmetry of the original Lagrangian.
In our case, we take $\beta=\frac{N_{f}}{32\pi^{2}}$.
$\phi\phi^{*}$ in eq.(\ref{1}) is not calculable and one may
expect other  complicated forms.
Here we consider the simplest example for convenience.

The gauge group of the low energy theory is $SU(N_{c}-N_{f})$.
What we concern is the auxiliary part of this Lagrangian:
\begin{equation}
  \label{2}
  L_{AUX}=\frac{\beta g^{2}\lambda\lambda}{v}F_{\phi}+h.c.+
    F^{*}_{\phi}F_{\phi}    
\end{equation}
(This term can be derived directly by 1-loop calculation.)
We can simply assume that the cut-off scale of this effective
Lagrangian is $v$.
The factor of $g^{2}$ appears because we have rescaled gaugino fields
to have canonical kinetic terms.
The equation of motion for $F_{\phi}$ is:
\begin{eqnarray}
  \label{2a}
  \frac{\partial L}{\partial F_{\phi}}&=&\frac{\beta g^{2}}{v}\lambda\lambda
  +F_{\phi}^{*}\nonumber\\
  &=&0
\end{eqnarray}
This equation means that $<\lambda\lambda>$ is proportional to
$F_{\phi}$ so we can think that $<\lambda\lambda>$ is
the order parameter for the supersymmetry breaking.
Using the tadpole method\cite{tad} we can derive a gap equation directly
from (\ref{2a}).
\begin{eqnarray}
  \label{2b}
  F^{*}_{\phi}&\times&\left(1-4G^{2}\int \frac{d^{4}p}{(2\pi)^{4}}
   \frac{1}{p^{2}
    +m_{\lambda}^{2}}\right)=0\nonumber\\
  &&\left\{
    \begin{array}{l}
      G^{2}=\frac{\beta^{2} g^{4}}{v^{2}}(N_{c}-N_{f})\\
      m_{\lambda}^{2}=\frac{|F_{\phi}|^{2}g^{4}\beta^{2}}{v^{2}}        
\end{array}
  \right.
\end{eqnarray}
Of course one can derive (\ref{2b}) by explicit calculation
of 1-loop effective potential.
Let us examine the solution.
After integration we can rewrite it in a simple form.
\begin{equation}
  \frac{4\pi^{2}}{G^{2}\Lambda^{2}}=1-\left(\frac{m_{\lambda}^{2}}
  {\Lambda^{2}}\right)ln\left(1+\frac{\Lambda^{2}}{m_{\lambda}^{2}}
  \right)
\end{equation}
(See also Fig.1a.)
In the strong coupling region, this equation can have non-trivial
solution.
The explicit form of the potential is shown in Fig.1b.
(Here we ignore the trivial solution $F_{\phi}=0$ because such
a condensation-vanishing solution does exist also in the
 effective (composite)
 Lagrangian analysis
of pure Yang-Mills theory, but it is usually neglected.)
Let us examine the behavior of this non-trivial
solution.
In pure supersymmetric Yang-Mills theories, gaugino condensation
is observed even in the weak coupling region because of the
instanton calculation
and Witten index argument that suggests
the invariance of Witten index in the deformation of 
coupling constants\cite{witten}.
If we believe that the characteristics of the low energy
Lagrangian of massless SQCD is 
also similar to pure SYM, the weak coupling region should be
lifted by gaugino condensation effect.
On the other hand, if we believe that non-compactness of the moduli
space is crucial 
and believe that gaugino condensation
should vanish in the weak coupling region, we can think that the potential 
represented in Fig.1c is reliable and potential is flat in the weak coupling
region.
We cannot make definite answer to this question, but some suggestive
arguments can be given by adding a small mass term to the field
$\phi$.
\begin{equation}
  \label{3.0}
  L^{add}_{mass}=\frac{1}{2}\epsilon \phi^{2}
\end{equation}
Existence of this term suggests that the moduli space is now compact.
The resulting gap equation is drastically changed.
We can naturally set $F$-components vanish, and the equation turns out
to be a non-trivial equation for ``$\phi$''.
Relevant terms are:
\begin{equation}
  \label{3}
  L_{AUX}=\left(\frac{\beta g^{2}}{v}\lambda\lambda+\epsilon 
  \phi\right)F^{*}_{\phi}
  +h.c.+F^{*}_{\phi}F_{\phi}
\end{equation}
The equation of motion for $F_{\phi}$ suggests that $<\lambda\lambda>$
is now proportional to $\phi$ and no longer an order parameter for
the supersymmetry breaking.
The gap equation is given by:
\begin{eqnarray}
  \label{3a}
  \epsilon \phi&\times&\left(
  1-4G^{2}
  \int \frac{d^{4}p}{(2\pi)^{4}}
  \frac{1}{p^{2}+m_{\lambda}^{2}}\right)=0\nonumber\\
    &&\left\{
    \begin{array}{l}
    G^{2}=\frac{\beta^{2} g^{4}}{v^{2}}(N_{c}-N_{f})\\
    m_{\lambda}^{2}=\frac{\epsilon^{2} g^{4}\beta^{2} |\phi|^{2}}{v^{2}}
    \end{array}
    \right.
\end{eqnarray}
In general, this equation has a solution $m_{\lambda}=const.$
(see Fig.2a and 2b) which 
does {\it not} break supersymmetry(Fig.2c), and does not 
change Witten index for any(non-zero) value of $\epsilon$ and $g_{0}$.
In this case, the potential energy is always $0$ for any value of $g$.
Because the moduli space is compact in this case, 
it is reliable that there is no phase
transition of gaugino condensation.

Now let us examine the limit $\epsilon\rightarrow 0$.
In our model 
$\phi$ does not run away to infinity.
If we take $\phi$  so large that the theory is weakly coupled,
then non-trivial solution disappears and only the solution $\phi=0$
is left.
This contradicts the assumption, so we think $\phi$
is finite and the potential is stabilized.
Taking $\epsilon\rightarrow 0$, we can find a solution
$4\pi^{2}/G^{2}\Lambda^{2}=1$ (see Fig.2d) i.e.:
\begin{eqnarray}
 G^{2}&=&\frac{\beta^{2}g(\phi)^{4}}{\Lambda^{2}}(N_{c}-N_{f})\nonumber\\
 &=&\frac{4\pi^{2}}{\Lambda^{2}}
\end{eqnarray}
This means that we can find a solution at $\phi=c\Lambda exp(-\frac{1}
{\beta g_{0}^{2}})$, here $c$ is a constant
($c=exp(\beta^{2}(N_{c}-N_{f})/
4\pi^{2})$).
Because we have fixed the symmetry breaking scale $v$ and
considered it as a cut-off scale for the low energy effective theory,
we cannot find a runaway solution for $<\overline{Q}Q>$ from
this low energy Lagrangian.

\section{Dilaton dependent coupling constant}
\hspace*{\parindent}
In this section we mainly focus on the supersymmetric
Yang-Mills theory in which the gauge coupling constant is dependent
only on the dilaton field $S$.
The Lagrangian is now written as:
\begin{equation}
  \label{4}
  L=\frac{1}{4} f(S) 
  W^{\alpha}W_{\alpha}+h.c.-\Lambda^{2}log(S+\overline{S})
\end{equation}
Here we assume $Ref(S)=ReS\equiv \frac{1}{g_{0}^{2}}$.
Relevant part of the Lagrangian is:
\begin{equation}
  \label{4a}
  L_{AUX}=g^{2}F_{S}\lambda\lambda+h.c.+\frac{F^{*}_{S}F_{S}}{(S+
    \overline{S})^{2}}\Lambda^{2}
\end{equation}
Here $g$ means the renormalized coupling constant.
Using the tadpole method we can find the following gap equation,
\begin{eqnarray}
  \label{4b}
  F_{S}^{*}&\times&\left(
  \frac{\Lambda^{2}}{(S+\overline{S})^{2}}-4G^{2}\int \frac{d^{4}p}{(2\pi)^{4}}
  \frac{1}{p^{2}+m_{\lambda}^{2}}\right)=0\nonumber\\
  &&\left\{
    \begin{array}{l}
      G^{2}=g^{4}N_{c}\\
      m_{\lambda}^{2}=g^{4}|F_{S}|^{2}
    \end{array}
  \right.
\end{eqnarray}
which can be rewritten as:
\begin{eqnarray}
  F_{S}^{*}&\times&\left(
  1-4G'^{2}\int \frac{d^{4}p}{(2\pi)^{4}}
  \frac{1}{p^{2}+m_{\lambda}^{2}}\right)=0\nonumber\\
&&\left\{
    \begin{array}{l}
      G^{2}=\frac{4}{\Lambda^{2}}\left(\frac{g}{g_{0}}\right)^{4}N_{c}\\
      m_{\lambda}^{2}=g^{4}|F_{S}|^{2}
    \end{array}
  \right.
\end{eqnarray}
This equation relies only on the parameter $g/g_{0}$.
This gap equation has the same characteristics of massless SQCD,
so the dilaton potential is flat in the weak coupling region.
The potential is almost the same as Fig.1c.

Then what would happen if we add a small mass term,
for example, $L_{m}=\epsilon\Lambda^{2} S^{2}/2$ ?
Now the auxiliary part of the Lagrangian is:
\begin{equation}
  L_{AUX}=\left[g^{2}F_{S}\lambda\lambda+
  \epsilon \Lambda^{2}S F_{S}+h.c.\right]+\frac{F^{*}_{S}F_{S}}{(S+
    \overline{S})^{2}}\Lambda^{2}
\end{equation}
This means that the dilaton field $S$ is now becomes an order parameter
for $<\lambda\lambda>$.

The gap equation is now given by:
\begin{eqnarray}
  \label{5}
  \epsilon S&\times&\left( 1-4G^{2}
  \int \frac{d^{4}p}{(2\pi)^{4}} \frac{1}{p^{2}+m^{2}_{\lambda}}
   \right)=0\nonumber\\
   &&\left\{
   \begin{array}{lll}
     G^{2}&=&\frac{g^{4}(S+\overline{S})^{2}}{\Lambda^{2}}N_{c}\\
          &=&\frac{4}{\Lambda^{2}}\left(\frac{g}{g_{0}}\right)^{4}N_{c}\\
     m_{\lambda}^{2}&=&\epsilon^{2} |S|^{2} g^{4}(S+\overline{S})^{4}\\
                    &=&16\epsilon^{2}\left(\frac{g}{g_{0}}\right)^{4}|S|^{4}
   \end{array}
   \right.
\end{eqnarray}
This equation has a non-trivial solution at finite value
once we fix $g/g_{0}$.

Let us comment on the phenomenological models.
If a small mass term is induced by the small spacetime curvature or some 
effects of higher theories, and assuming that our analyses are 
collect in such theories, we can expect that the analysis above 
may be used to analyze the dynamical dilaton potential and its runaway
problem.
It is important that we found a vacuum with finite $S$,
without introducing multiple gauge groups.

\section{Gaugino condensation in supergravity}
\hspace*{\parindent}  
In the standard superfield formalism of the locally supersymmetric action,
we have:
\begin{eqnarray}
  \label{lag}
  S&=&\frac{-3}{\kappa^{2}}\int d^{8}z E exp\left(-\frac{1}{3}
  \kappa^{2} K_{0} \right)\nonumber\\
  &&+\int d^{8}z{\cal E}\left[W_{0}+\frac{1}{4}f_{0}{\cal WW}\right]
    +h.c.
\end{eqnarray}
Here we set $\kappa^{2}=8\pi/M_{p}^{2}$.
In the usual formalism of minimal supergravity, the Weyl rescaling is done 
in terms of component fields.
However, in order to understand the anomalous quantum corrections 
to the classical action, we need a manifest supersymmetric formalism,
in which the Weyl rescaling is also supersymmetric.
It is easy to see that the classical action(\ref{lag}) itself is not
super-Weyl invariant.
However, the lack of the super-Weyl invariance can be  recovered
with the help of a chiral superfield $\varphi$(Weyl compensator).

For the classical action (\ref{lag}), the K\"ahler function $K_{0}$, the
superpotential $W_{0}$ and the gauge coupling $f_{0}$ are modified\cite{kap}:
\begin{eqnarray}
  \label{modify}
  K_{0} &\rightarrow& K=K_{0}-6\kappa^{-2}{\sf Re} log\varphi \nonumber\\
  W_{0} &\rightarrow& W=\varphi^{3}W_{0}\nonumber\\
  f_{0} &\rightarrow& f=f_{0}+\xi log\varphi
\end{eqnarray}
$\xi$ is the constant chosen to cancel the super-Weyl anomaly.
The super-Weyl transformations contain an R-symmetry in its
imaginary part, so we can think that this is a natural extension
of \cite{ross} in which
 a compensator for the R-symmetry played a crucial role.

Let us examine the simplest case.
We include an auxiliary field $H$  and 
set the form of $W_{0}$ and $f_{0}$ as:
\begin{eqnarray}
  \label{fix}
  W_{0}&=& \lambda H^{3}\nonumber\\
  Ref_{0}&=& \frac{1}{g^{2}_{0}}
\end{eqnarray}
and rescale the field $\varphi$ as:
\begin{equation}
  \tilde{\varphi}=H \varphi
\end{equation}
where H is some auxiliary field.
Finally we have:
\begin{eqnarray}
  \label{fin}
  K&=&K_{0}-6\kappa^{-2}{\sf Re}
  log\left(\frac{\tilde{\varphi}}{\Lambda}\right)
  \nonumber\\
  W&=&\lambda {\tilde{\varphi}}^{3}\nonumber\\
  f&=&\frac{1}{g^{2}_{0}}+\xi log\left(\frac{\tilde{\varphi}}{\Lambda}\right)
\end{eqnarray}
From the equation of motion for the auxirialy field of the super-Weyl
compensator, we have the relation:
\begin{equation}
  \label{aux0}
  \lambda \tilde{\varphi}^{3}-\frac{\xi}{6}
  g^{2}\lambda^{\alpha}\lambda_{\alpha}=0
\end{equation}
And the tree level scalar potential is:
\begin{eqnarray}
  \label{pot1}
  V_{0}&=&-3\kappa^{2}|W|^{2}\nonumber\\
  &=&-3\kappa^{2}\lambda^{2}|\tilde{\varphi}^{3}|^{2}
\end{eqnarray}
The equation of motion for the auxirialy field(\ref{aux0})
 suggests that eq.(\ref{pot1}) can be interpreted as a four-fermion
interaction of  the gaugino:
\begin{equation}
  \label{4f}
  -\frac{1}{12}\kappa^{2}g^{4}
  \xi^{2}|\lambda^{\alpha}\lambda_{\alpha}|^{2}
\end{equation}
This four-fermion interaction becomes strong as $\frac{1}{g^{2}}={\sf Re} f$ 
reaches 0.
The strong coupling point is:
\begin{equation}
  \label{strong}
  \tilde{\varphi}_{s}=\Lambda e^{-\frac{1}{g^{2}_{0}\xi}}
\end{equation}
Using the tadpole method we can have a gap equation:
\begin{eqnarray}
  \label{g}
  \lambda \tilde{\varphi}^{3}&\times&
  \left(1-4G^{2}
    \int \frac{d^{4}p}{(2\pi)^{4}}
    \frac{1}{p^{2}+m_{\lambda}^{2}}\right)=0\nonumber\\
  &&\left\{
    \begin{array}{l}
      G^{2}=\frac{\xi^{2}\kappa^{2}g^{4}N_{c}}{12}\\
      m_{\lambda}^{2}=\frac{\kappa^{4}\xi^{2}g^{4}\lambda^{2}
        |\tilde{\varphi}^{3}|^{2}}{4}
    \end{array}
  \right.
\end{eqnarray}
The solution for the gap equation(\ref{g}) is given in Fig.3a.
We can see that there is always a solution for non-zero 
gaugino condensation.
(We have implicitly assumed that the vacuum expectation value for the
auxiliary field for the Weyl compensator superfield is $0$.)
For a second example, we include the dilaton superfield $S$.
Now $f_{0}$ is not a constant and depends on the field $S$:
\begin{equation}
  f_{0}=S
\end{equation}
And the K\"ahler potential for the dilaton superfield is:
\begin{equation}
  K_{0}=-\kappa^{-2}log(S+\overline{S})
\end{equation}
Here we should include the effect of the 
 dilaton field in the scalar potential.
The tree level scalar potential is:
\begin{equation}
  V_{0}=h_{S}(G^{-1})^{S}_{S}h^{S}-3\kappa^{2}|W|^{2}
\end{equation}
The auxirialy field for $S$ is:
\begin{eqnarray}
  \label{aux}
  h_{S}&=&\kappa^{2}\left[\frac{1}{2} \frac{W}{S+\overline{S}}
  +\frac{1}{4}f_{S}
  \lambda^{\alpha}\lambda^{\alpha}\right]\nonumber\\
  &=&\frac{\kappa^{2}}{4}W\frac{1+12S_{R}\xi^{-1}}{S_{R}}
\end{eqnarray}
Here we set $G=K+ln(\frac{1}{4}|W|^{2})$ and $S_{R}=(S+\overline{S})/2$.
The tree level potential can be given in a simple form
\begin{equation}
  \label{sc}
  V_{0}=\lambda^{2}A|\tilde{\varphi}^{3}|^{2}
\end{equation}
where
\begin{equation}
  A=\frac{1}{16}
  \kappa^{2}\left[\left(1+\frac{12S_{R}}{\xi}\right)^{2}
  -3\right].
\end{equation}
In this case, the gap equation is given by:
\begin{eqnarray}
  \label{gg}
  \lambda \tilde{\varphi}^{3}&\times&
  \left(1-4G^{2}
    \int \frac{d^{4}p}{(2\pi)^{4}}
    \frac{1}{p^{2}+m_{\lambda}^{2}}\right)=0\nonumber\\
  &&\left\{
    \begin{array}{l}
      G^{2}=\frac{\lambda^{2}\xi^{2}g^{4}N_{c}A}{36}\\
      m_{\lambda}^{2}=\frac{A^{2}\xi^{2}
        \lambda^{2}|\tilde{\varphi}^{3}|^{2}}{36}
    \end{array}
  \right.
\end{eqnarray}
Now let us consider the difference between our result and ref.\cite{ross}.
In ref.\cite{ross}, the solution for the gap equation is estimated
after fixing the coupling constant at $g_{c}$ which is introduced
by hand.
It is true that the effective potential is singular at $\tilde{\varphi}_{s}$
(\ref{strong}),
but without introducing the cut-off, we can find a solution for (\ref{gg})
at finite value.(see Fig 3a)

Another important point is the stability of the dilaton potential.
It is suggested in \cite{ross} that the dilaton potential
has a stable vacuum without introducing multiple gauge groups
if we use the Nambu-Jona-Lasinio method.
Related topics are also discussed in \cite{add}.
Can we explain this phenomenon along the line of  the previous 
section?
The scalar potential (\ref{sc}) contains a dilaton bilinear 
that can be interpreted as a mass term.
As is shown in section 3, such a mass term would stabilize the
dilaton potential.

\section{Conclusion}
\hspace*{\parindent}
We examined the formation of gaugino condensation in the hidden
sector within a Nambu-Jona-Lasinio type approach.
First we considered  global supersymmetric gauge theories
that have the $SU(N_{c})$ gauge group and $N_{f}$ matter fields.
We can find the phase transition in massless SQCD, but
in the massive theory,  we cannot
find such a phase transition. 
We can conclude that 
gaugino condensation is always non-zero in massive SQCD.
 
We also examined a model with a dilaton dependent coupling constant.
The result is similar to SQCD.
We found a stabilized dilaton potential 
when  we add a small mass term.
We also extended our analysis to  supergravity.

\section*{Acknowledgment}
\hspace*{\parindent}
We thank K.Fujikawa, T.Hotta and K.Tobe for many helpful discussions.

\end{document}